\documentclass[aps,article,nofootinbib,twocolumn,groupedaddress]{revtex4}
\usepackage{graphicx}
\usepackage{amsmath,amssymb}
\usepackage{hyperref}
\usepackage{bm}
\usepackage{color}   

\hypersetup{
    colorlinks=true,
    linkcolor=red,
    citecolor=blue,
}

\def\be{\begin{equation}}
\def\ee{\end{equation}}
\def\ba{\begin{eqnarray}}
\def\ea{\end{eqnarray}}

\newcommand{\bn}{\hat{\bf n}}

\begin{document}

\title{CMB Faraday rotation as seen through the Milky Way}

\author{Soma De$^{1}$, Levon Pogosian$^{2,3}$, Tanmay Vachaspati$^{4}$}

\affiliation{$^1$School of Earth and Space Exploration, Arizona State University, Tempe, AZ 85287, USA  \\
$^2$Department of Physics, Simon Fraser University, Burnaby, BC, V5A 1S6, Canada \\
$^3$Centre for Theoretical Cosmology, DAMTP, University of Cambridge,
CB3 0WA, UK \\
$^4$Physics Department, Arizona State University, Tempe, AZ 85287, USA  
}

\begin{abstract}
Faraday Rotation (FR) of CMB polarization, as measured through mode-coupling correlations of E and B modes, can be a promising probe of a stochastic primordial magnetic field (PMF). While the existence of a PMF is still hypothetical, there will certainly be a contribution to CMB FR from the magnetic field of the Milky Way. We use existing estimates of the Milky Way rotation measure (RM) to forecast its detectability with upcoming and future CMB experiments. We find that the galactic RM will not be seen in polarization measurements by Planck, but that it will need to be accounted for by CMB experiments capable of detecting the weak lensing contribution to the B-mode. We then discuss prospects for constraining the PMF in the presence of FR due to the galaxy under various assumptions that include partial de-lensing and partial subtraction of the galactic FR. We find that a realistic future sub-orbital experiment, covering a patch of the sky near the galactic poles, can detect a scale-invariant PMF of 0.1 nano-Gauss at better than 95\% confidence level, while a dedicated space-based experiment can detect even smaller fields.
\end{abstract}
\maketitle

\section{Introduction}

The discovery of a primordial magnetic field\footnote{In the context of this paper, ``primordial'' means that the magnetic field was generated prior to last scattering.} (PMF) would have a profound impact on our understanding of the early universe \cite{Grasso:2000wj} and would help explain the origin of the observed magnetic fields in galaxies and clusters \cite{Widrow:2002ud}. Several observational probes are currently being investigated and the cosmic microwave background (CMB) is a promising tool to discover and study the PMF on the largest cosmic scales.

A PMF can affect the thermal distribution and the polarization of the CMB. The most relevant CMB signature depends on the form of the magnetic field spectrum and also on the level of instrumental noise for different observations. For instance, current CMB bounds of a few nG \cite{Ade:2013lta,Paoletti:2008ck,Paoletti:2012bb} on a scale-invariant PMF \cite{Turner:1987bw,Ratra:1991bn} derive from temperature anisotropies 
sourced by the magnetic stress-energy. Comparable bounds were recently obtained in \cite{Kahniashvili:2012dy} using the Lyman-$\alpha$ forest spectra \cite{Croft:2000hs}. As the stress-energy is quadratic in the magnetic field strength, improving this bound by a factor of $2$ would require a $16$ fold improvement in the accuracy of the spectra. On the other hand, Faraday Rotation (FR) of CMB polarization, being linear in the magnetic field strength, offers an alternative way to improve CMB bounds on a scale-invariant PMF by an order of magnitude \cite{Yadav:2012uz}, and possibly more, with the next generation CMB experiments. Even current CMB data is close to providing competitive bounds on
scale-invariant PMF through their FR signature \cite{Kahniashvili:2008hx,Kahniashvili:2012dy}.

In contrast, a PMF produced causally in an aftermath of the electroweak or QCD 
phase transition \cite{Vachaspati:1991nm} would have a blue spectrum 
\cite{Durrer:2003ja,Jedamzik:2010cy} with most of the power concentrated near a
 small cutoff scale set by the plasma conductivity at recombination. It has been
suggested \cite{Jedamzik:2011cu} that such small-scale fields can appreciably alter 
the recombination history via enhancement of small scale baryonic inhomogeneities.
 Consequently, the strongest CMB constraint would come from an overall shift in the
distance to last scattering. According to \cite{Jedamzik:2011cu}, upcoming CMB
experiments can rule out causally produced fields with a comoving strength larger 
than $10^{-11}~{\rm G}$.

Given the promise of FR to significantly improve the bounds on scale-invariant PMF
 \cite{Yadav:2012uz}, it is important to assess the strength of the rotation induced by
the magnetic fields in our own galaxy. A rotation measure (RM) map of the Milky Way 
has recently been assembled by Oppermann et al \cite{Oppermann:2011td} based on 
an extensive catalog of FR of compact extragalactic polarized radio sources. They also 
presented the angular power spectrum of RM and find that it is very close to a scale-
invariant spectrum at $\ell \lesssim 200$. Fig.~1 of \cite{Oppermann:2011td} shows that the 
root-mean-square RM away from the galactic plane is about $20$-$30~{\rm rad/m}^2$.
Additional insight into galactic RM on smaller scales can be 
gained from Haverkorn et al \cite{Haverkorn:2003ad} who studied Stokes parameters 
generated by diffuse polarized sources residing inside the Milky Way. Their power 
spectra generally agree with the scale-invariance of the large scale tail of the power 
spectrum of \cite{Oppermann:2011td}, but also show evidence of a break in the 
spectrum, indicating a power law suppression of the RM power on small scales. 
Such a break in the RM spectrum is also present in the model of Minter and Spangler \cite{MinterSpangler1996} who, by fitting to a small set of RM data from extragalactic sources, argued that the spectrum should be set by Kolmogorov turbulence on small scales.

The approximately scale-invariant shape of the galactic RM spectrum can obscure FR 
constraints on scale-invariant PMF. Based on calculations in \cite{Pogosian:2011qv,Yadav:2012uz}, 
one can estimate that a typical galactic RM of $30~{\rm rad/m}^2$ corresponds to an {\it effective}, or ``energy equivalent'' (see Eq.~(\ref{Beff}) for the definition) PMF strength of $0.6$ nG. This is well below the bounds from Planck \cite{Ade:2013lta} obtained using non-FR diagnostics and, as we confirm in this paper, below the levels detectable by Planck via the mode-coupling statistics induced by FR.  
However, as shown in \cite{Yadav:2012uz}, future CMB experiments capable of detecting the weak lensing B-mode will be able to constrain a scale-invariant PMF of $0.1$ nG strength at $100$ GHz. Operating at lower frequencies and combining FR information from several channels may further improve the constraints. Clearly, with such high sensitivities the contribution of the galactic RM to the total FR signal will become important.

In this paper, we investigate the imprint of the galactic RM on CMB observables, and its impact on detectability of the PMF via the EB and TB mode-coupling correlations. We start by introducing the necessary concepts and reviewing the known galactic RM measurements in Sec.~\ref{FRCMB}. In Sec.~\ref{sec:detection}, we estimate detectability of the galactic RM by upcoming and future CMB experiments and forecast future bounds on the scale-invariant PMF under various assumptions. We conclude with a summary in Sec.~\ref{sec:summary}.

\section{Faraday Rotation of CMB polarization}
\label{FRCMB}

\subsection{Basics of Faraday Rotation}
\label{basics}

A CMB experiment measures Stokes parameters in different directions on the sky, 
with parameters $Q$ and $U$ quantifying linear polarization. If CMB photons pass 
through ionized regions permeated by magnetic fields, the direction of linear 
polarization is rotated by an angle \cite{Kosowsky:1996yc,Harari:1996ac} 
\begin{equation}
\alpha(\hat{\bf n}) = \lambda_0^2 \ RM(\hat{\bf n})= \frac{3}{{16 \pi^2 e}} \lambda_0^2 
\int \dot{\tau} \ {\bf B} \cdot d{\bf l} \ ,
\label{alpha-FR}
\end{equation}
where $\hat{\bf n}$ is the direction along the line of sight, $\dot{\tau}$ is the differential optical depth, 
$\lambda_0$ is the observed wavelength of the radiation, ${\bf B}$ is the ``comoving'' magnetic field strength (the physical field strength scales with the expansion as ${\bf B}^{\rm phys}={\bf B}/a^2$) and $d{\bf l}$ is the comoving length element along the photon trajectory. The rotation measure, $RM$, is a frequency independent quantity used to describe the strength of FR. Under the rotation of the polarization vector, the two Stokes parameters transform as
\begin{equation}
Q(\nu) + iU(\nu) =(Q^{(0)} + iU^{(0)}) \exp(2i\alpha(\nu)) \ ,
\label{qu-rotation}
\ee
where $Q^{(0)}$ and $U^{(0)}$ are the Stokes parameters at last scattering. As
an approximation, $Q^{(0)}$ and $U^{(0)}$ can be taken to be the observed 
Stokes parameters at a very high frequency since the FR falls off as $1/\nu^2$.

A PMF contributes to FR primarily at the time of last scattering, just after the polarization was generated, while the mean ionized fraction was still high, and
the field strength was strongest. 
Subsequently, additional FR is produced in ionized regions along the line of sight that contain magnetic fields, such as clusters of galaxies and our own galaxy. Eq.~(\ref{alpha-FR}) implies that a significant FR angle can be produced by a small magnetic field over a very large distance, which is the case at recombination, or by a larger magnetic field over a smaller path, which is the case for the Milky Way. We will not discuss FR from clusters which are likely to have a white noise spectrum and contribute to CMB polarization on small scales. We are more concerned with the contribution from the Milky Way that can look very similar to that of a scale-invariant PMF.

In theory, it is possible to extract a map of the FR angle by taking maps of 
$Q$ and $U$ at different frequencies and using Eq.~(\ref{qu-rotation}) 
to solve for the rotation in each pixel.
Each additional frequency channel provides a separate measurement of $\alpha(\nu)$ 
thus reducing the error bar on the measurement of $RM(\bn)$. Such a direct 
measurement of FR may be challenging when the $Q$ and $U$ signal in each pixel is 
dominated by noise. 

Another way to extract the rotation field is from correlations between E and B modes induced by FR using quadratic estimators \cite{Kamionkowski:2008fp,Yadav:2009eb,Gluscevic:2009mm,Gluscevic:2012me,Yadav:2012tn}, which is analogous to the method introduced in \cite{Hu:2001kj} for isolating the weak lensing contribution to CMB anisotropy. Unlike a direct extraction of FR from Eq.~(\ref{qu-rotation}), the quadratic estimator method does not utilize frequency dependence and is statistical in nature. It formally involves summing over all pixels of $Q$ and $U$ in order to reconstruct $\alpha$ in a given direction on the sky. 

For small rotation angles, the relation between the spherical expansion coefficients of the E, B and $\alpha$ fields can be written as \cite{Gluscevic:2009mm}
\be
B_{lm}=2\sum_{LM}\sum_{l' m'}\alpha_{LM} E_{l' m'} 
\xi_{lml'm'}^{LM}H_{ll'}^L \ ,
\label{eq:blm}
\ee
where $\xi_{lml'm'}^{LM}$ and $H_{ll'}^L$ are defined in 
terms of Wigner $3$-$j$ symbols as \cite{Gluscevic:2009mm}
\ba
\nonumber
\xi_{lml'm'}^{LM} &\equiv& (-1)^m \sqrt{ (2l+1)(2L+1)(2l'+1) \over 4\pi} \\
&\times& \left(
\begin{array}{ccc}
l  & L  & l'  \\
-m  & M  & m'    
\end{array}
\right) \\
H_{ll'}^L &\equiv& 
\left(
\begin{array}{ccc}
l  & L  & l'  \\
2  & 0  & -2    
\end{array}
\right) \ ,
\ea
and the summation is restricted to even $L+l'+l$. Eq.~(\ref{eq:blm}) implies 
correlations between multipoles of E and B modes that are caused by the FR. 
Since the primordial T and E are correlated, FR also correlates T and B. 

\begin{figure}[tbp]
\includegraphics[height=0.46\textwidth]{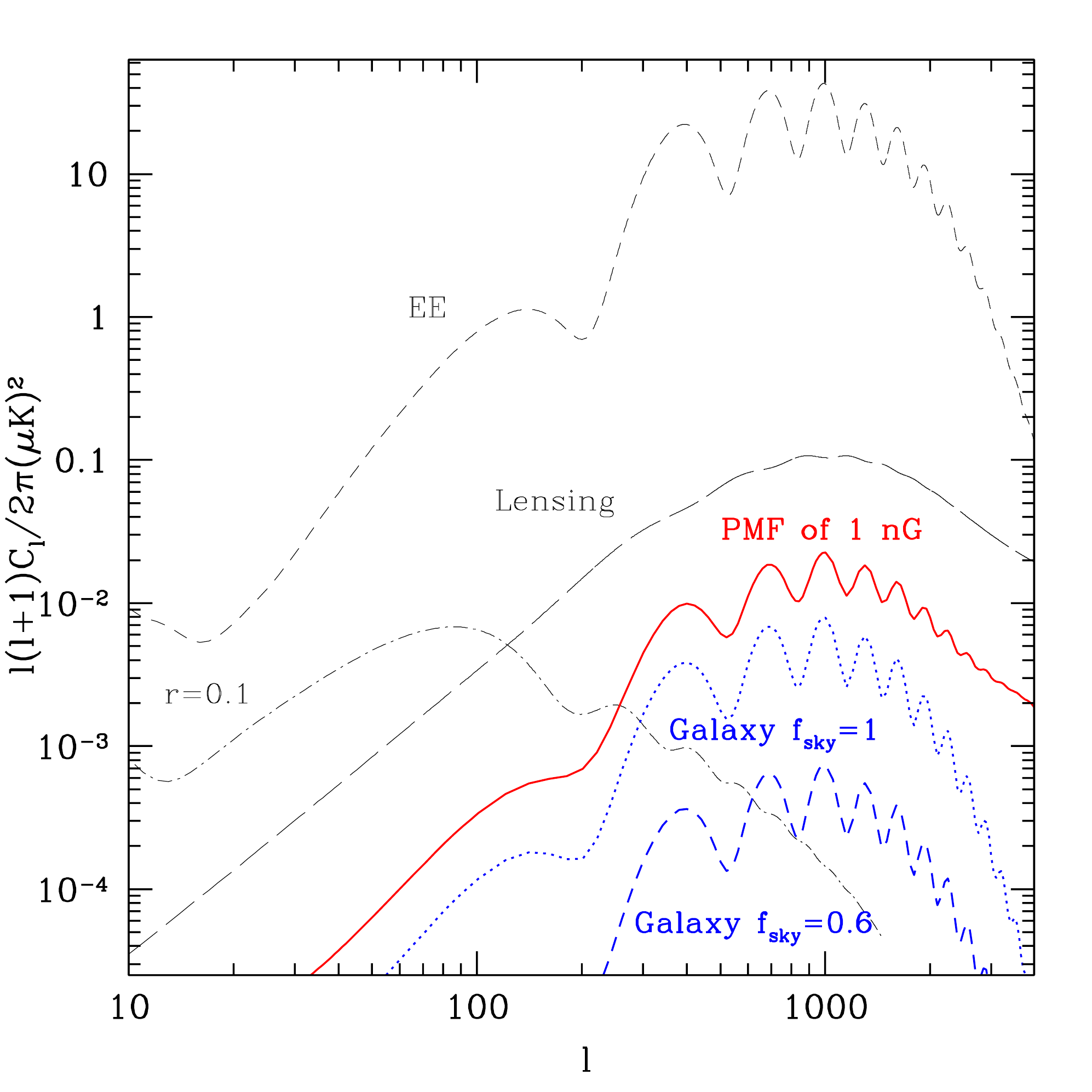}
\caption{The CMB B-mode spectrum from Faraday rotation sourced at $30$~GHz 
by a scale-invariant primordial magnetic field of 1 nG strength (red solid), by the full sky 
galactic magnetic field (blue dot), and by the galactic field with Planck's sky 
mask ($f_{\rm sky}=0.6$). The black short-dash line is the input E-mode spectrum, 
the black dash-dot line 
is the contribution from inflationary gravitational waves with $r=0.1$, while the black 
long-dash line is the expected contribution from gravitational lensing by large scale 
structure.}
\label{fig:cl}
\end{figure}

The quadratic estimator method is based on the assumption of statistical isotropy of primordial perturbations. This implies statistical independence of primordial $E_{l m}$ and $B_{l m}$ for different $lm$ pairs, {\it e.~g.} $\langle E_{l m}^* E_{l'm'} \rangle =  \delta_{ll'} \delta_{mm'} C_l^{EE}$ for the primordial E-mode. Furthermore, if primordial fields are Gaussian, all of their correlation functions can be expressed in terms of the power spectra. FR introduces correlations between unequal $lm$ and generates connected four-point correlations. The corresponding non-Gaussian signal can be used to extract the rotation angle. Namely, given a CMB polarization map, one constructs quantities such as \cite{Kamionkowski:2008fp,Gluscevic:2009mm}
\be
{\hat D}_{ll'}^{LM,{\rm map}}={4\pi \over (2l+1)(2l'+1)} \sum_{mm'} B_{lm}^{\rm map}E_{l'm'}^{\rm map*} \xi_{lml'm'}^{LM}
\label{eq:dll}
\ee
which is the minimum variance unbiased estimator for $D_{ll'}^{LM} = 2 \alpha_{LM}C_l^{EE}H_{ll'}^L$ as shown  in \cite{Pullen:2007tu}. Then, each pair of $l$ and $l'$ provides an estimate of $\alpha_{LM}$ via
\be
[{\hat \alpha}_{LM}]_{ll'} = {{\hat D}_{ll'}^{LM,{\rm map}} \over 2C_l^{EE}H_{ll'}^L} \ .
\label{alphallpr}
\ee
The minimum variance estimator ${\hat \alpha}_{LM}$ is obtained from an appropriately weighted sum
over estimators for each $ll'$ pair. If all such pairs were statistically independent, the weighting would
be given by the inverse variance of each estimator. However, one has to account for the correlation 
between the $ll'$ and $l'l$ pairs and a detailed derivation of the appropriate weighting can be found in \cite{Kamionkowski:2008fp}.
Quantities such as ${\hat D}_{ll'}^{LM,{\rm map}}$ can also be constructed from products of T and B or, more generally,
all possible quadratic combinations, {\it i.e.} \{EB, BE, TB, BT, TE, ET, EE\}. One can construct an estimator
${\hat \alpha}_{LM}$ that utilizes all these combinations while accounting for the covariance between them \cite{Gluscevic:2009mm}.
In this paper, we opt to consider EB/BE and TB/BT separately, since one of them is typically 
much more informative than other combinations. For further details of the method the reader is referred to 
\cite{Kamionkowski:2008fp,Gluscevic:2009mm}.

We will assume a ``stochastic'' primordial magnetic field such that its value at any given location
is described by a probability distribution function. This implies that one can only 
predict statistical properties of FR, such as the rotation power spectrum, 
$C_L^{\alpha \alpha}$. The estimator for the power spectrum directly follows 
from the estimator of $\alpha_{LM}$ and derives from four-point correlations, 
such as EBEB. In the next section, we will examine detectability of the rotation 
spectrum caused by the galactic and primordial magnetic fields.

We note that the form of the quadratic estimator allows for contributions from the 
monopole ($L=0$) and dipole ($L=1$) of the FR field which, in principle, should not be 
ignored. However, the monopole is generally not expected for FR, since it would imply 
a non-zero magnetic charge enclosed by the CMB surface, while the dipole, 
corresponding to a uniform magnetic field, is strongly constrained for the PMF and is 
relatively unimportant for the galactic RM.

In addition to mode-coupling correlations of EB and TB type, FR also contributes 
to the B-mode polarization spectrum, $C_l^{BB}$. In Fig.~\ref{fig:cl} we show the B-
mode spectrum at $30$ GHz due to FR by a PMF with a scale-invariant spectrum and 
effective field strength of $1~{\rm nG}$.The B-mode 
spectrum due to FR by the galactic field is also shown for two choices of sky cuts.  For 
reference, we show the $E$ mode auto-correlation spectrum which acts as a source 
for the FR B modes, the B modes from inflationary gravitational waves with $r=0.1$, 
and the expected contribution from weak lensing (WL). As can be seen from Fig.~
\ref{fig:cl}, the shape of the FR induced B mode spectrum largely mimics that of the E-
mode, and the galactic contribution has a very similar shape to that of the scale-
invariant PMF. A more detailed discussion of the FR induced B-mode spectrum can be 
found in \cite{Pogosian:2011qv}. As shown in Sec.~\ref{sec:detection}, the 
FR signatures of the galactic or scale-invariant PMF are always less visible in 
the B-mode spectrum 
compared to the EB quadratic estimator, with the latter promising to provide the 
tightest constraints. We note that for causally generated PMF, which have blue spectra 
with most of the power concentrated at the dissipation scale, the signal-to-noise ratio 
(SNR) for detecting FR in BB is higher than that in the quadratic estimators 
\cite{Yadav:2012uz}, although the resulting constraints are not competitive with 
other probes of 
causal fields. In what follows, we will focus on FR signatures of scale-invariant PMF 
only.

\subsection{A scale-invariant primordial magnetic field}
\label{pmf}

Assuming statistical homogeneity and isotropy, the 
magnetic field correlation function in Fourier space is~\cite{1975mit..bookR....M}
\begin{eqnarray}
\langle {b}^* _i ({\bm k}) {b} _j ({\bm k}') \rangle
&=&
\left[ \left(\delta _{ij} - {k_i k_j \over k^2} \right) 
 S (k)  + i \epsilon_{ijl} {k_l \over k} A(k) \right]
\nonumber \\
&& \hskip 2 cm \times (2 \pi)^3 \delta^3 ({\bm k} -{\bm k}') 
\end{eqnarray}
where repeated indices are summed over, $b_i({\bm k})$ denotes the Fourier transform of the magnetic field at
wave vector ${\bm k}$, $k=|{\bm k}|$, and $S(k)$ and $A(k)$ are the symmetric and 
antisymmetric (helical) parts of the magnetic field power spectrum. 

The helical part of the power spectrum does not play a role in Faraday Rotation; 
only the symmetric part of the spectrum is relevant for us. On scales larger than
the inertial scale, {\it i.e.} $k < k_I$, $S(k)$ will fall off as a power law. On scales 
smaller than a dissipative scale, {\it i.e.} $k > k_d$, the spectrum will get sharply
cut-off. A second power-law behavior is possible for $k_I < k < k_d$. 
However, we shall assume that $k_I \approx k_d$, so that the magnetic field
spectrum is given by a single power law behavior at all scales larger than the 
dissipation scale. These features can be summarized by \cite{Pogosian:2011qv}
\begin{eqnarray}
S(k) = \begin{cases}
\Omega_{B\gamma} {\rho}_\gamma 
             \frac{32\pi^3 n}{k_I^3} 
   \left ( \frac{k}{k_I} \right ) ^{2n-3}, &\mbox{$0<k<k_{\rm diss}$}\\
0, & \mbox{$k_{\rm diss} < k$}
       \end{cases}
\label{eq:PB}
\end{eqnarray}
where $\Omega_{B\gamma}$ is the ratio of cosmological magnetic and photon 
energy densities, and $\rho_\gamma$ is the photon energy density.

For convenience, we define an ``effective'', or ``energy equivalent'',
magnetic field strength
\begin{equation}
B_{\rm eff} \equiv ( 8\pi \Omega_{B\gamma} \rho_\gamma )^{1/2}
 = 3.25\times 10^{-6}\sqrt{\Omega_{B\gamma}} ~{\rm Gauss}.
\label{Beff}
\end{equation}
A uniform magnetic field of strength $B_{\rm eff}$ will have the same
energy density as a stochastic magnetic field with spectrum in
Eq.~(\ref{eq:PB}).

Scale invariance for $k < k_d$ corresponds to $n=0$, in which case
the expressions in Eq.~(\ref{eq:PB}) are not well defined. One reason is
that the energy density diverges logarithmically for a strictly scale invariant
field. We get around this issue by only considering approximately scale
invariant magnetic fields for which $n$ is small.  Note that, for scale invariant fields, $B_{\rm eff}$ is equal
to the field strength smoothed over a scale $\lambda$ (if it is larger than the dissipation scale), or $B_\lambda$, frequently used in the literature ({\it e.g.} in \cite{Ade:2013lta,Paoletti:2008ck,Paoletti:2012bb}).

As we have already noted, the galactic RM spectrum is approximately scale 
invariant. The RM induced by a scale invariant PMF is also scale invariant 
\cite{Yadav:2012uz}. Hence the amplitudes of the galactic and PMF rotation measure 
spectra can be related directly for a scale invariant PMF.

For a scale-invariant angular spectrum $C_\ell$, the quantity $\ell(\ell+1)C_\ell/2\pi$ is constant. Hence, one can define an amplitude
\begin{equation}
A_{\rm RM} = \sqrt{\frac{L(L+1)C_L^{\rm RM} }{2\pi}}
\end{equation}
for the range of $L$ over which the RM spectrum is approximately scale-invariant. One can see from Fig.~\ref{fig:rmcl} that this holds for the galactic $RM$ for $L\lesssim 100$. Similarly, one can define an amplitude 
\be
A_\alpha = \sqrt{L(L+1) C^{\alpha \alpha}_L \over 2\pi}
\ee
The relation between the Faraday rotation angle in radians and the RM is 
given by $\alpha = c^2 \nu^{-2} {\rm RM} =  10^{-4}  \nu^{-2}_{30} {\rm RM} ~{\rm m^2}$
where $\nu_{30}\equiv \nu/30~{\rm GHz}$.
For example, a characteristic galactic RM of $30~{\rm rad/m^2}$ at 
$30~{\rm GHz}$ gives a rotation angle of $3 \times 10^{-3}$ rad, or 
$0.17^\circ$. The amplitudes of $A_\alpha$ in radians and $A_{\rm RM}$ in ${\rm rad/m^2}$ are therefore also related via 
\be
A_\alpha= 10^{-4}  \nu^{-2}_{30} A_{\rm RM}~{\rm m^2} \ .
\label{alphaRM}
\ee

To gain some intuition into the relative importance of the galactic RM prior to more detail forecasts 
in Sec~\ref{detect-pmf}, let us estimate the $B_{\rm eff}$ of scale-invariant PMF that gives RM 
comparable to that of our galaxy. For this, we can compare the amplitude 
$A_\alpha$ from PMF to the amplitude for galactic spectra at the same frequency. The relation 
between PMF and $A_\alpha$ involves integration along the line of sight (Eq.~(\ref{alpha-FR})). 
Rather than trying to find an approximate analytical solution, we use exact spectra found numerically 
in \cite{Pogosian:2011qv,Yadav:2012uz}. From Fig.~1 of Ref.~\cite{Yadav:2012uz}, one can infer 
that for a scale invariant PMF with 
$\Omega_{B\gamma}=10^{-4}$, $(A^{\rm PMF}_\alpha)^2 \approx 2.43 \times 10^{-2}$ rad$^2$ at 
$30~{\rm GHz}$. We can then use Eqs.~(\ref{alphaRM}) and Eq.~(\ref{Beff}) to relate
$B_{\rm eff}$ to $A_{\rm RM}$:
\be
B_{\rm eff} \approx 0.021 \ A_{\rm RM} \ {\rm nG \ m^2 \over rad} \ .
\label{eq:BeffRM}
\ee
From Fig.~\ref{fig:rmcl}, we can see that $A_{\rm RM} \approx 30~{\rm rad/m^2}$ for the
 galactic RM spectrum obtained without any sky cuts. It follows from Eq.~(\ref{eq:BeffRM})
that this corresponds to a scale-invariant PMF with $B_{\rm eff} \approx 0.6~{\rm nG}$. Such 
PMF are well below levels detectable by Planck through their FR 
signal, but can be detected at high significance by future sub-orbital
and space-based experiments \cite{Yadav:2012uz}.

We can also estimate the peak value of the B-mode spectrum, $P_B \equiv \ell_{\rm peak}(\ell_{\rm peak}+1)C_{\ell_{\rm peak}}^{BB}/2\pi$, in ($\mu$K)$^2$, for a given $A_{\rm RM}$ and frequency.
From Fig.~\ref{fig:cl}, we see that $P_B \approx 0.03~(\mu$K)$^2 $ for a scale invariant PMF of $1$~nG strength. Using
this fact along with Eq.~(\ref{eq:BeffRM}), we can deduce  
\be
P_B  \approx {0.03 \over \nu^{4}_{30}}  \left[ {B_{\rm eff} \over {\rm nG}} \right]^2 (\mu K)^2
\approx  {10^{-5} \over  \nu^{4}_{30}}  \left[ {A_{\rm RM} \ {\rm m}^2 \over {\rm rad}} \right]^2 (\mu K)^2 \ .
\ee
Thus, for a typical galactic $A_{\rm RM}$ of $30$ rad/m$^2$, at $30$~GHz, 
we should expect a peak in the B-mode spectrum of about $10^{-2}$ ($\mu$K)$^2$, which
agrees with the blue dotted line in Fig.~\ref{fig:cl} corresponding to the unmasked galactic RM.
In practice, the galactic plane is masked in all CMB experiments, {\it e.g.} the mask applied to
Planck's $30$~GHz map only uses $0.6$ of the sky. This significantly reduces the amplitude of
the galactic RM spectrum and, as a consequence, the B-mode spectrum. This is shown as a 
blue dashed line in Fig.~\ref{fig:cl}. One can see that the galactic B-mode is at least one, and 
more likely two,
orders of magnitude below the weak lensing signal, so there is almost no 
hope of seeing it even at a frequency as low as $30$~GHz. Going a factor of $\sim 3$ 
higher in frequency, which is more typical in CMB studies, results in another factor of $\sim 100$ 
decrease in the B-mode spectrum.

One of the conclusions of this section is that the galactic RM is practically invisible 
in the BB auto-correlation. Hence, we should look at the mode-coupling estimators,
as they are known to be much more sensitive to FR \cite{Yadav:2012uz}. 
As shown in \cite{Yadav:2012uz}, future experiments can easily probe 
$0.1~{\rm nG}$ strength PMF fields with the mode-coupling 
estimators.

\subsection{The galactic rotation measure}
\label{galaxy}

\begin{figure}[tbp]
\includegraphics[height=0.85\textwidth]{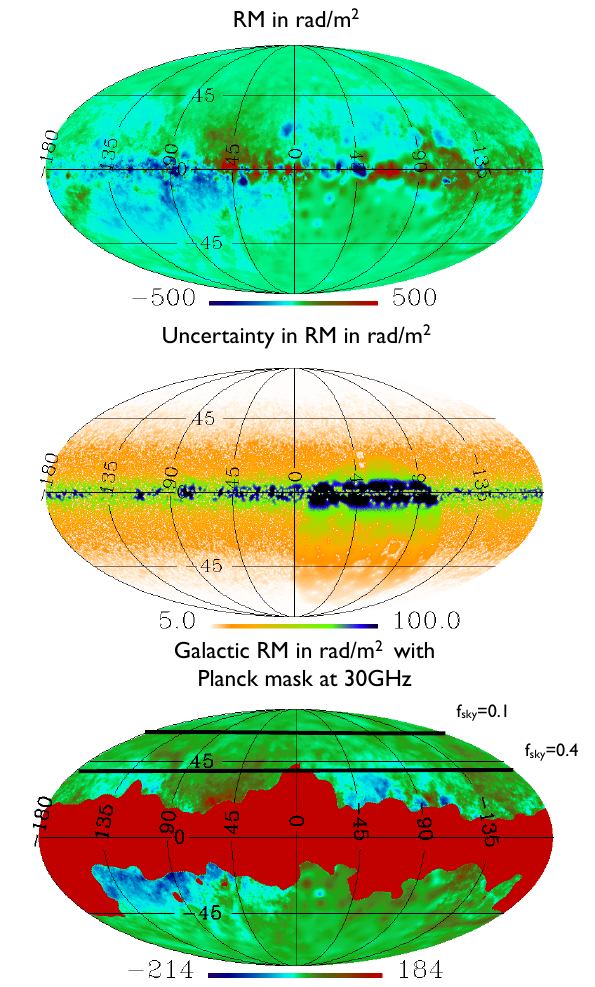}
\caption{Maps of RM from Oppermann et al \cite{Oppermann:2011td}. The top 
panel shows the full sky map, the middle panel shows the associated error, while the 
bottom panel shows the map after applying the Planck $30~{\rm GHz}$ sky cut.
The bottom panel also shows lines of symmetric cuts corresponding to 
$f_{\rm sky}=0.4$ and $f_{\rm sky}=0.1$. 
}
\label{fig:rmmap}
\end{figure}

\begin{figure}[tbp]
\includegraphics[height=0.4\textwidth]{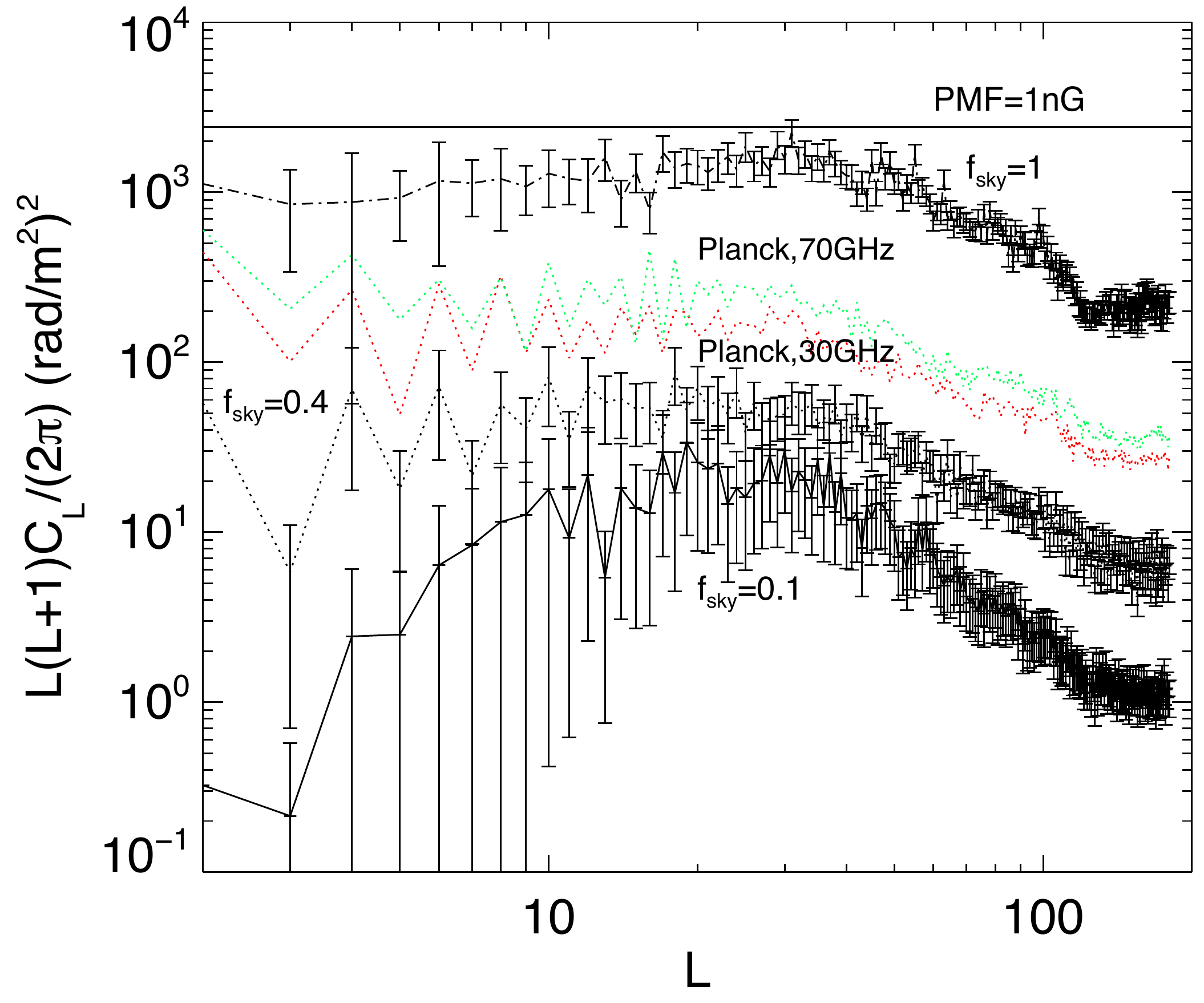}
\caption{
The RM angular spectra, $L(L+1)C_L^{\rm RM}/2\pi$, obtained from the 
RM map of Oppermann et al \cite{Oppermann:2011td} with different cuts. 
Shown are the 
RM spectra corresponding to, from top to bottom, a scale-invariant PMF of $1$ nG, 
galaxy with no sky cut, with a mask used by Planck for their $70$ GHz map, a Planck 
mask for the $30$ GHz  map, and symmetric cuts corresponding to $f_{\rm sky}=0.4$
and $f_{\rm sky}=0.1$. 
}
\label{fig:rmcl}
\end{figure}

A rotation measure (RM) map of Milky Way has been put together by 
Oppermann et al \cite{Oppermann:2011td} based on an extensive catalog of Faraday 
rotation data of compact extragalactic polarized radio sources. The RM is strongest 
along the galactic plane, but remains significant at the poles. However, the latitude 
dependence of the rms RM disappears at latitudes higher than about $80^\circ$, 
where the dependence flattens out at a rms value of about $10$ rad/m$^2$. 
Oppermann et al \cite{Oppermann:2011td} also presented the angular power spectrum of RM calculated after factoring out the latitude dependence. They find that $L(L+1)C^{\rm RM}_L \propto L^{-0.17}$ at $L \lesssim 200$, which is very close to a scale-invariant spectrum.

Some insight into galactic RM on smaller scales is provided by Haverkorn et al
\cite{Haverkorn:2003ad}. They studied Stokes parameters, generated by diffuse 
polarized sources residing inside the galaxy, at several frequencies around 350 MHz. 
They focused on two separate patches away from the Galactic plane, each covering 
about $50$ sq. deg. The rotation measures in both were less than $10$ rad/m$^2$. 
However, since they are looking at sources inside the galaxy, they only see a fraction 
of the total RM along each line of sight. Thus, their numbers cannot be directly 
compared to RM of \cite{Oppermann:2011td}. Still their power spectrum 
gives a sense for 
the scale dependence of the total RM, and generally agrees with the flat nature of the 
large scale tail of the power spectrum of \cite{Oppermann:2011td}.  At a scale of about 
$0.3^\circ$ ($L \sim 600$), they see evidence of a break in the spectrum, indicating a 
power law suppression of the RM power on small scales. Such a break in the spectrum 
is also present in the model of Minter and Spangler \cite{MinterSpangler1996}, whose
analysis is based on a small set of RM data from extragalactic sources. They 
argued that on small 
scales the spectrum should be due to Kolmogorov turbulence. The corresponding 
scaling of $L(L+1)C_L^{\rm RM}$ would be $L^{-2/3}$. To summarize, based on the evidence in 
\cite{Oppermann:2011td,Haverkorn:2003ad,MinterSpangler1996}, if we were to 
measure RM in a few degree patches near galactic poles, we would expect a RM 
spectrum that is roughly scale invariant at $L<600$ turning into a $L^{-2/3}$ tail on 
smaller scales. As we will show in the next section, the quadratic estimator of the FR angle is largely uninformative for $L>300$. 
Hence, we do 
not need a RM map with a resolution  higher than that of Oppermann et al 
\cite{Oppermann:2011td}.

Note that the CMB polarization even in a small patch around the poles is affected by 
very long wavelength modes of variations of the RM. 
In terms of Eq.~(\ref{eq:blm}), lowest multipoles of $\alpha_{LM}$ contribute 
significantly to the much higher multipoles of $B_{l m}$. Thus, to predict the 
observability of the FR in CMB in a small patch one cannot simply use the high $\ell$ 
portion of the RM spectrum. 
We should use the full RM spectrum obtained from the RM after applying a 
mask normally used to block the galactic plane.

In Fig.~\ref{fig:rmmap} we show the RM map for our galaxy obtained from the Faraday rotation of extragalactic
sources in units of rad/m$^{2}$ by Oppermann et al \cite{Oppermann:2011td}. The top panel is the full map. The mid-panel shows the uncertainty in RM in rad/m$^{2}$. Most of the signal and uncertainty come from the
galactic disk. The bottom panel shows the RM map with part of the sky removed with a mask used for Planck's mask $30$ GHz map \cite{planck:maps}. The bottom panel also shows lines of symmetric latitude cuts corresponding to sky fractions of $f_{\rm sky}=0.1$ and $0.4$.

In Fig.~\ref{fig:rmcl} we present the power spectra $L(L+1)C^{\rm RM}_{L}/2\pi$ of the galactic RM under different symmetric sky cuts and two Planck mask cuts: one for $30$ GHz with $f_{\rm sky}=0.6$, and one for $70$ GHz with $f_{\rm sky}=0.7$. The figure also shows the RM spectrum from a scale-invariant PMF with $B_{\rm eff}=1$ nG. To calculate the galactic RM spectra, we first use HEALPix~\cite{healpix,Gorski:2004by} to find
\be
{\tilde a}_{LM} = \int d\Omega \ W(\Omega) \ R(\Omega) \ Y^*_{LM}(\Omega),
\label{tildealm}
\ee
where $R(\Omega)$ is the RM field, and $W(\Omega)$ is the mask which is $0$ if a pixel is masked and $1$ if it is not. We then evaluate
\be
C^{\rm RM}_L = {1 \over f_{\rm sky}(2L+1)} \sum_{M=-L}^{L} {\tilde a}^*_{LM} {\tilde a}_{LM} \ ,
\label{eq:clrm}
\ee
which are the so-called ``pseudo-$C_L$'' rescaled by $f_{\rm sky}$.

The commonly used procedure for estimating $C_L$'s from a partial sky is detailed in \cite{Hivon:2001jp}. It assumes statistical isotropy and involves calculating a conversion matrix that relates ``pseudo-$C_L$'', obtained from a partial sky,  to  the ``full-sky'' $C_L$. In Eq.~(\ref{eq:clrm}), as a crude approximation, we ignore the mode-coupling and replace the matrix with a rescaling by the $f_{\rm sky}$ factor. Note that the galactic RM map is not isotropic and the ``full-sky'' spectrum reconstructed from a given patch is not the same as the actual spectrum evaluated from the full-sky. Thus, our estimates in Sec.~\ref{sec:detection} of the detectability of the galactic RM spectrum, and of the detectability of the PMF with the galactic RM as a foreground, are specific to particular masks.  The error bars in Fig.~\ref{fig:rmcl} receive contributions from the uncertainty in the measurements of the RM map and from the cosmic variance.

\section{Detectability prospects}
\label{sec:detection}

We are interested in answering two questions: 1){\it How well can CMB polarization experiments 
measure the galactic RM spectrum?}  2){\it What bounds on a scale-invariant PMF can be placed from 
measurements of FR?} Both contributions,  the galactic FR and the primordial FR, contribute in the 
same way to the quadratic estimators discussed in Sec.~\ref{basics}.  Let ${\hat \alpha}^{EB}_{LM}$ 
be the estimator of the rotation angle that one reconstructs from linear combinations of quadratic 
terms $E^*_{lm}B_{l'm'}$. The variance in ${\hat \alpha}^{EB}_{LM}$, for a statistically
isotropic rotation angle, is defined as 
\begin{equation}
\langle {\hat \alpha}^{EB*}_{LM} {\hat \alpha}^{EB}_{L'M'} \rangle =\delta_{LL'} \delta_{MM'} 
\sigma^2_L \ , 
\label{variance-alm}
\end{equation}
with $\sigma^2_L$ given by \cite{Kamionkowski:2008fp,Gluscevic:2009mm}
\be
\sigma^2_L = C_L^{\alpha \alpha} + N_L^{EB} \ ,
\label{eq:variance}
\ee
where $C_L^{\alpha \alpha}$ is the rotation power spectrum, and $N_L^{EB}$ is the part of the
variance that does not depend on the rotation angle and contains contributions from variances of individual
estimators ${\hat D}_{ll'}^{LM,{\rm map}}$ (see Eq.~(\ref{eq:dll})), 
while accounting for their covariance. It is given by \cite{Kamionkowski:2008fp,Gluscevic:2009mm}
\be
(N_L^{EB})^{-1} = {1\over \pi} \sum_{\ell \ell'} { (2\ell+1)(2\ell'+1)[C^{EE}_{\ell'} 
                         H^L_{\ell \ell'}]^2 \over {\tilde C}_{\ell'}^{EE} {\tilde C}_{\ell}^{BB}}  \ ,
\label{eq:ebnoise}
\ee
where $\ell+\ell'+L=$even, and
\be
{\tilde C}^{XX}_\ell \equiv [C^{XX}]^{\rm prim}_\ell+f_{\rm DL}[C^{XX}]^{\rm WL}_\ell+I^{XX}_\ell \ ,
\label{clvariance}
\ee
is the measured spectrum, with $XX$ standing for either EE or BB, that includes the primordial contribution, $[C^{XX}]^{\rm prim}$, the weak lensing contribution, $[C^{XX}]^{\rm WL}$, and the noise term associated with the experiment, $I^{XX}_\ell$. We allow for the possibility of partial subtraction of the weak lensing contribution by introducing a de-lensing fraction, $f_{\rm DL}$, in (\ref{clvariance}).
The efficiency of de-lensing depends on the method used
to de-lens \cite{Hirata:2002jy,Seljak:2003pn}, as well as the noise and the resolution parameters of the experiment. 
In \cite{Seljak:2003pn} it was found that the quadratic estimator method of de-lensing
\cite{Hu:2001kj}, 
which is similar but orthogonal \cite{Kamionkowski:2008fp,Yadav:2009eb} to the one 
we use to extract the FR rotation, can reduce the lensing contribution to BB by up to a 
factor of $7$ as it reaches the white noise limit. On the other hand, the so-called 
iterative method \cite{Seljak:2003pn} can further reduce the lensing contribution
by a factor of $10$ or larger. In our analysis we have only considered two cases -- one with no de-lensing ($f_{\rm DL}=1$) 
and one with a factor of $100$ reduction of the lensing B-mode ($f_{\rm DL}=0.01$). 
We adopt the latter optimistic assumption with the aim of
illustrating the relative significance of lensing contamination for an experiment of given 
noise and resolution. Note that the numerator in Eq.~(\ref{eq:ebnoise}) contains the 
unlensed E-mode spectrum, while the denominator includes the lensing contribution to 
E- and B-mode spectra. Predictions for both are readily obtained from CAMB \cite{camb,Lewis:1999bs} for a 
given cosmological model. 

The instrumental noise term, $I^{XX}_\ell$, accounts for the detector noise as well as the suppression of the spectrum on scales smaller than the width of the beam \cite{Bond:1997wr}:
\be
I^{EE}_\ell=I^{BB}_\ell= \Delta^2_P \exp(\ell^2 \Theta^2_{\rm FWHM} / 8\ln 2) \ ,
\label{eq:polnoise}
\ee
where $\Delta_P$ quantifies detector noise associated with measurements of CMB polarization and $\Theta_{\rm FWHM}$ is the full width at half maximum of the Gaussian beam.

In the case of TB, the expression for the variance analogous to Eq.~(\ref{eq:ebnoise}) is \cite{Gluscevic:2009mm}
\be
(N_L^{TB})^{-1} = {1\over \pi} \sum_{\ell \ell'} { (2\ell+1)(2\ell'+1)[C^{TE}_{\ell'} 
H^L_{\ell \ell'}]^2 \over {\tilde C}_{\ell'}^{TT} {\tilde C}_{\ell}^{BB}} .
\label{eq:tbnoise}
\ee

Eq.~(\ref{variance-alm}) assumes that the distribution of the rotation angle is isotropic, and the angular brackets denote an ensemble average over many realizations of $\alpha$ as well as realizations of the primordial CMB sky. This is a well-motivated assumption for the PMF and for the galactic RM near galactic poles. Away from the poles, the galactic RM shows a strong dependence on the galactic latitude \cite{Oppermann:2011td}. This means that the forecasted variance can differ from the actual uncertainty for larger values of $f_{\rm sky}$, although we expect it to be of the correct order of magnitude. For the galactic RM map, we take the rotation angular spectrum to be $C^{\alpha \alpha}_L \equiv \nu^{-4} C_L^{RM}$, with $C_L^{RM}$ calculated from the RM map of \cite{Oppermann:2011td} after applying sky cuts, as plotted in Fig.~\ref{fig:rmcl} and explained in Sec.~\ref{galaxy}. 

We will consider several sets of experimental parameters corresponding to ongoing, future and hypothetical experiments. Namely, we consider Planck's $30$ GHz LFI and $100$ GHz HFI channels based on actual performance characteristics \cite{Ade:2013ktc}, POLARBEAR \cite{Keating:2011iq} at $90$ GHz with parameters compiled in \cite{Ma:2010yb},  QUIET Phase II \cite{Samtleben:2008rb} at $40$ GHz using the parameters compiled in \cite{Ma:2010yb},  $30$, $45$, $70$ and $100$ GHz channels of a proposed CMBPol satellite \cite{Baumann:2008aq}, as well as an optimistic hypothetical sub-orbital and space-based experiments at $30$ and $90$ GHz. 
The assumed sky coverage ($f_{\rm sky}$), resolution ($\Theta_{\rm FWHM}$), 
and instrumental noise ($\Delta_P$) parameters are listed in 
Tables~\ref{table:galaxy} and \ref{table:pmf}.

\subsection{Detectability of the galactic rotation measure}

\begin{figure}[tbp]
\includegraphics[height=0.45\textwidth]{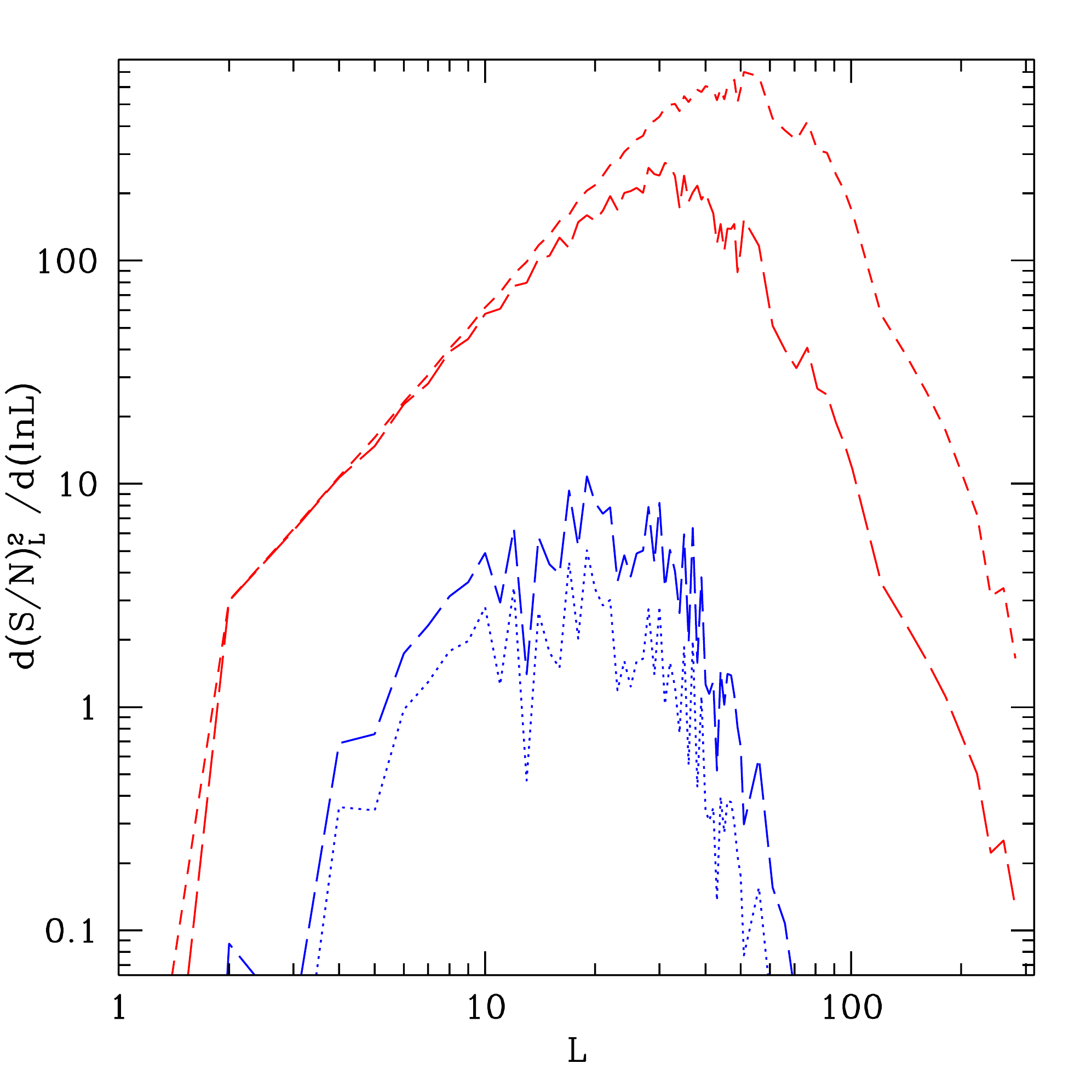}
\caption{Contribution of individual multipoles to the net SNR of detection of the 
galactic RM spectrum. Plotted are $d(S/N)_L^2 / d \ln L$ for an optimistic $30$ GHz sub-orbital 
experiment with (solid blue) and without (blue dot) de-lensing, as well as for a hypothetical future $30$ 
GHz space probe with (red short dash) and without (red long dash) de-lensing 
($f_{\rm DL}=0.01$). The spiky nature of the plot is due to the spikiness of the RM spectra, as seen in Fig.~\ref{fig:rmcl}. Experimental parameters assumed in this plot are given in the text and in Table \ref{table:galaxy}.}
\label{fig:dsngal}
\end{figure}

\begin{table*}[tbp] 
\centering      
\begin{tabular}{c c c c c c c}  
\hline\hline                        
Name - freq (GHz) & $f_{\rm sky}$ & FWHM (arcmin) &  $\Delta_P$($\mu$K-arcmin) & $(S/N)_{EB}$ (+DL) & $(S/N)_{TB}$ (+DL) & $(S/N)_{BB}$ (+DL) \\ 
\hline                    
Planck LFI - 30 & 0.6 &  33  & 240  &  5.3E-4 (same) & 2.2E-3 (same)  & 2.3E-4 (same)       \\ 
Planck HFI - 100 & 0.7 &  9.7  & 106  &  1.4E-3 (same) & 7.5E-4 (same)  & 6E-5 (same)       \\ 
Polarbear - 90 & 0.024$^a$ &  6.7  & 7.6  &  1.3E-2 (1.5E-2) & 1.6E-3 (2.0E-3)  & 4.6E-4 (6.0E-4)    \\
QUIET II - 40  & 0.04$^a$ &  23  & 1.7  &  0.3 (0.8) & 0.05 (0.2)  & 0.02 (0.08)    \\
CMBPOL - 30 & 0.6 &  26  & 19  &  1.0 (same) & 0.4 (same)  & 0.05 (same) \\
CMBPOL - 45 & 0.7 &  17  & 8.25  &  2.1 (2.3) & 0.8 (0.9)  & 0.12 (0.15) \\
CMBPOL - 70 & 0.7 &  11  & 4.23  &  2.0 (2.6) & 0.6 (0.9)  & 0.08 (0.14) \\
CMBPOL - 100 & 0.7 &  8  & 3.22  &  1.4 (2.0) & 0.3 (0.6)  & 0.03 (0.07) \\
Suborbital - 30 & 0.1 & 1.3  & 3  &  2.0 (3.1) & 0.3 (0.7)  &  0.08 (0.2) \\
Space - 30 & 0.6 & 4  & 1.4  &  18 (28) & 7 (14)  &  5 (30) \\
Space - 90 & 0.7 & 4  & 1.4  &  3.3 (6.8) & 1.0 (2.4)  &  0.09 (0.64) \\
\hline     
\end{tabular} 
\caption{S/N of the overall detection of the galactic RM spectrum with Planck, Polarbear, QUIET, CMBPOL and optimistic future sub-orbital and space experiments. Results are presented without and with (+DL) de-lensing by a factor $f_{\rm DL}=0.01$. ($^a$ based on 0.1 of RM sky.)} 
\label{table:galaxy}  
\end{table*}

Let us estimate the significance level at which the galactic RM spectrum can be detected. The signal in this case is the rotation angle spectrum $C_L^{\alpha \alpha,G}$, related to the galactic RM spectrum via $C_L^{\alpha \alpha,G} = C_L^{\rm RM,G}/\nu^4$. Our forecasts will be specific to the spectra estimated after applying particular masks to the RM map from \cite{Oppermann:2011td}, and the RM spectra are different for different sky cuts, as can be seen in Fig.~\ref{fig:rmcl}. The approximate expression for the variance is obtained under the assumption of statistically isotropic and Gaussian rotation angle:
\be
\sigma^2_{C_L} = {2 \sigma^2_L \over f_{\rm sky}(2L+1)} 
\ee
where $\sigma^2_L$ is the variance of the estimator ${\hat \alpha}^{XB}_{LM}$ defined in Eq.~(\ref{eq:variance}), and $X$ stands for either E or T, depending on whether one uses the EB or TB estimator. The total SNR squared of the detection of the galactic FR spectrum is the sum of $(C_L^{\alpha \alpha,G/\sigma_{C_L})^2}$ at each $L$:
\be
\left( S \over N \right)^2_{XB} = \sum_{L=1}^{L_{max}} {f_{\rm sky} (2L+1) [C_L^{\alpha \alpha,G}]^2 \over 2 [C_L^{\alpha \alpha,G}+N^{XB}_L]^2} \ .
\label{eq:ebsnrG}
\ee
Note that even for a small sky coverage, the SNR receives a non-zero contribution from the smallest 
multipoles $L$ of the rotation angle. This is because large angle features of RM couple small angle 
features of CMB, and there will be $ll'$ pairs (see Eq.~(\ref{alphallpr})) giving an estimate of RM at 
small $L$ no matter how small $f_{\rm sky}$ is. However, the smaller the sky coverage, the smaller 
is the number of available $ll'$ pairs leading to larger statistical errors.

In addition to the EB estimator, one may also want to know how detectable 
the Milky Way RM is in the B-mode spectrum. In this case, the signal is the B-mode spectrum
generated by the FR inside the galaxy, $C_L^{BB,G}$ (see Fig.~\ref{fig:cl}). The squared SNR is
\be
\left( S \over N \right)^2_{BB} = \sum_{L=2}^{L_{max}} {f_{\rm sky} \over 2}
(2L+1)  \left( C_L^{BB,G} \over {\tilde C}^{BB}_L \right)^2 \ .
\label{eq:bbsnr}
\ee
where ${\tilde C}^{BB}_L$ is given by Eq.~(\ref{clvariance}) and includes contributions from galaxy, weak lensing 
and instrumental noise.

In Fig.~\ref{fig:dsngal}, we plot contributions to the net SNR in detection of the galactic 
RM spectrum, given by Eq.~(\ref{fig:dsngal}), received  per $\ln L$. We show four different 
cases, all at $30$ GHz, corresponding to hypothetical future sub-orbital and space 
based experiments, with and without de-lensing by a factor $f_{\rm DL}=0.01$. We 
assume that the sub-orbital experiment will cover $f_{\rm sky}=0.1$ with $\Delta_P=3 
\mu$K-arcmin and FWHM of $1.3'$, while the space-based probe will cover $f_{\rm 
sky}=0.6$ (based on Planck's $30$ GHz sky mask) with $\Delta_P=1.4 \mu$K-arcmin 
and FWHM of $4'$. Note that, as shown in Fig.~\ref{fig:rmcl}, elimination of the 
galactic plane significantly reduces the amplitude of the galactic RM spectrum signal. 
Thus, the overall SNR of detection of the galactic RM spectrum is only of ${\cal O}(1)$ for the most 
optimistic sub-orbital experiment, with de-lensing making a relatively minor difference.
In contrast, a space-based probe can detect the galactic RM at a high confidence 
level, with most of the signal coming from $4<L<100$. This means that CMB 
polarization can, in principle, be used to reconstruct the galactic RM map at a 
resolution of up to a degree. De-lensing the CMB maps can further improve the 
accuracy of the reconstruction.

In Table~\ref{table:galaxy}, we forecast the SNR in detection of the galactic RM spectrum for several ongoing, proposed and hypothetical experiments. For experiments with $f_{\rm sky}<0.1$, such as QUIET and POLARBEAR, we use $C_L^{RM}$ with the $f_{\rm sky}=0.1$ cut of the RM map (the bottom line in Fig.~\ref{fig:rmcl}) around the galactic plane. Thus, our estimates assume that QUIET and POLARBEAR will observe in patches that are close to the galactic poles where statistical properties of the galactic RM become independent of the exact size of the patch. This expectation is justified by the fact that latitude dependence essentially disappears as one approaches the poles, which is quantified in Fig.~1 of \cite{Oppermann:2011td}.

We separately show results based on the EB, TB and BB estimators, with and without de-lensing. The results allow us to make the following conclusions. Firstly, the galactic RM will be invisible in Planck's polarization maps. Secondly, the galactic RM spectrum is unlikely to be detected by a sub-orbital experiment covering a small fraction of the sky near the galactic poles, at least not via the mode-coupling quadratic estimators or its contribution to the B-mode spectrum, unless very optimistic assumptions are made about its resolution and sensitivity. Thirdly, space-based polarization probes, such as the proposed CMBPOL mission, should take the galactic RM into account.
We note that our analysis does not cover the possibility of the galactic RM being detected directly from Eq.~(\ref{qu-rotation}) by utilizing the frequency dependence of Stokes parameters. This may turn out to be a more sensitive method for future multi-frequency CMB experiments with sufficiently low instrumental noise. We leave investigation of this possibility for future work.

\subsection{Detectability of the primordial magnetic field}
\label{detect-pmf}

\begin{figure}
\includegraphics[height=0.45\textwidth]{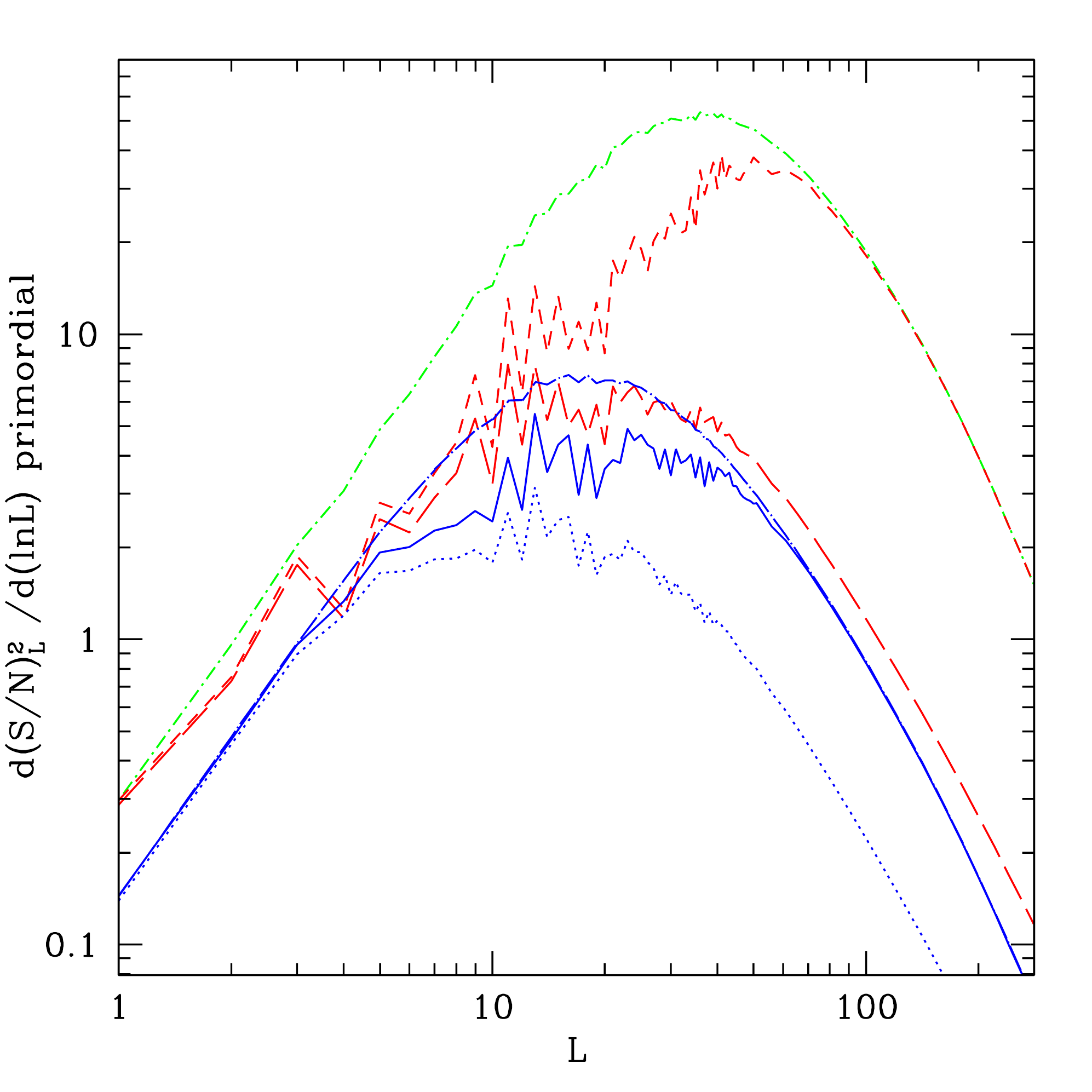}
\caption{Contribution of individual multipoles to the net SNR of detection of the PMF of $0.1$ nG. Plotted are $d(S/N)_L^2 / d \ln L$ for an optimistic $30$ GHz sub-orbital experiment without de-lensing (blue dot), after de-lensing ($f_{\rm DL}=0.01$, solid blue) and after de-lensing and partially subtracting the galactic RM (($f_{\rm DG}=0.1$, blue long dash-dot), as well as for a hypothetical future $30$ GHz space probe with (red short dash) and without (red long dash) de-lensing ($f_{\rm DL}=0.01$), as well as with partial ($f_{\rm DG}=0.1$) subtraction of the galactic RM (green dash-dot). The red short and long dash lines correspond to $f^{\rm opt}_{\rm sky}=0.2$, while the green dash-dot uses $f^{\rm opt}_{\rm sky}=f_{\rm sky}=0.5$. The lines are spiky because the RM spectra, as seen in Fig.~\ref{fig:rmcl}, contribute to then the noise part of the SNR. Note that spikes disappear when the galactic RM is partially subtracted.
Experimental parameters assumed in this plot are given the text and in Table \ref{table:pmf}.}
\label{fig:dsnpmg}
\end{figure}

\begin{table*}
\centering      
\begin{tabular}{c c c c c c c}  
\hline\hline                        
Name - freq (GHz) & $f_{\rm sky}$ ($f^{\rm opt}_{\rm sky}$) &  FWHM (arcmin) &  $\Delta_P$($\mu$K-arcmin) &  $B_{\rm eff}$ (2$\sigma$, nG)  & +DL (nG)  & +DL+DG (nG)  \\ 
\hline                    
Planck LFI - 30 & 0.6 &  33  & 240  & 16$^b$  & same & same   \\ 
Planck HFI - 100 & 0.7 &  9.7  & 106   & 23 & same & same  \\ 
Polarbear - 90 & 0.024$^a$ &  6.7  & 7.6   & 3.3 & 3.0 & same  \\
QUIET II - 40  & 0.04$^a$ &  23  & 1.7  & 0.46 & 0.26 & 0.25 \\ 
CMBPOL - 30 & 0.6 &  26  & 19  &  0.56 & 0.55 & 0.51 \\
CMBPOL - 45 & 0.7 &  17  & 8.25  &  0.38 & 0.35 &  0.29 \\
CMBPOL - 70 & 0.7 &  11  & 4.23  &  0.39 & 0.32 & 0.26 \\
CMBPOL - 100 & 0.7 &  8  & 3.22  &  0.52 & 0.4 & 0.34 \\
Suborbital - 30 & 0.1 & 1.3  & 3  &  0.09 & 0.07 & 0.05 \\
Suborbital - 90 & 0.1 & 1.3  & 3  &  0.63 & 0.45 & same \\
Space - 30 & 0.6 (0.2) & 4  & 1.4  &  0.06 & 0.04 & 0.02 \\
Space - 90 & 0.7 (0.4) & 4  & 1.4   & 0.26 & 0.15 & 0.12 \\
\hline     
\end{tabular} 
\caption{The expected $2\sigma$ bound in nano-Gauss on the strength of a 
scale-invariant PMF. Without de-lensing and with de-lensing (+DL) by a factor 
$f_{\rm DL}=0.01$, and with additional removal of the galactic RM by a factor 
$f_{\rm DG}=0.1$ (+DL+DG). Note that for full sky experiments, there is an optimal 
sky cut ($f^{\rm opt}_{\rm sky}$) that gives the best bounds on the PMF. ($^a$ 
based on 0.1 of RM sky; $^b$ from TB estimator.)} 
\label{table:pmf}  
\end{table*}

\begin{figure*}
\begin{center}
\begin{tabular}{c c c}
\includegraphics[scale=0.25]{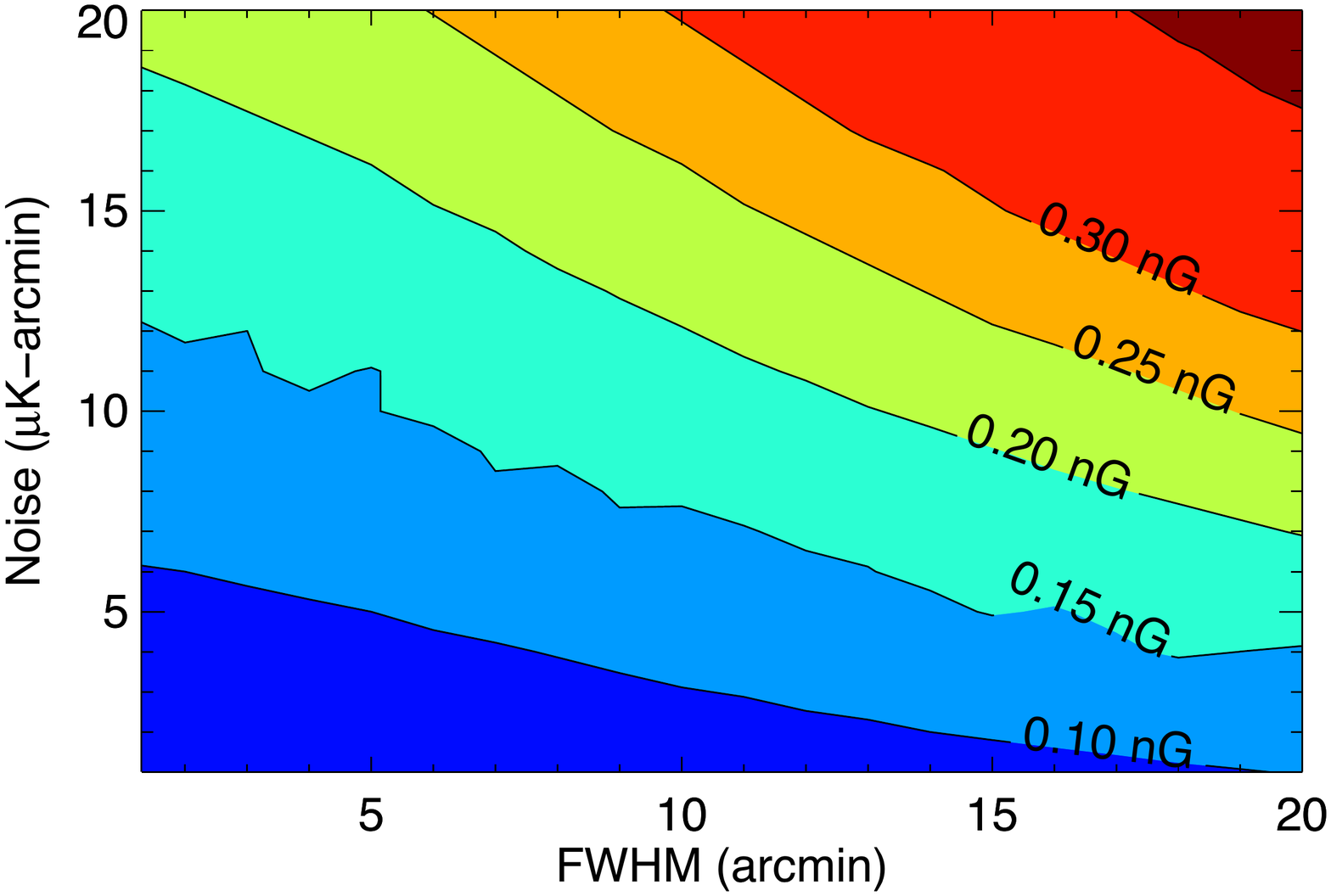}
\hskip -1 cm
\includegraphics[scale=0.25]{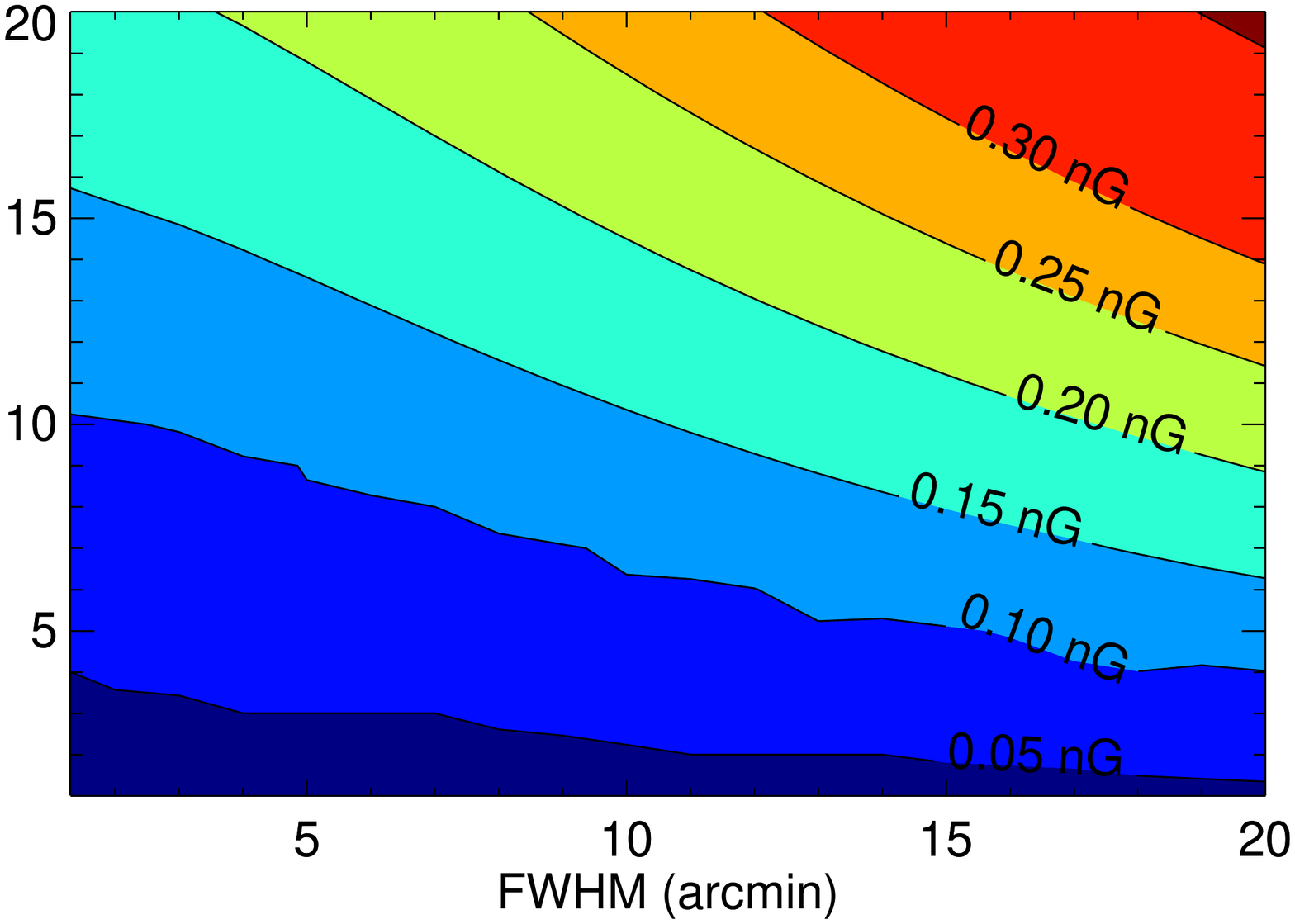}
\hskip -1 cm
\includegraphics[scale=0.25]{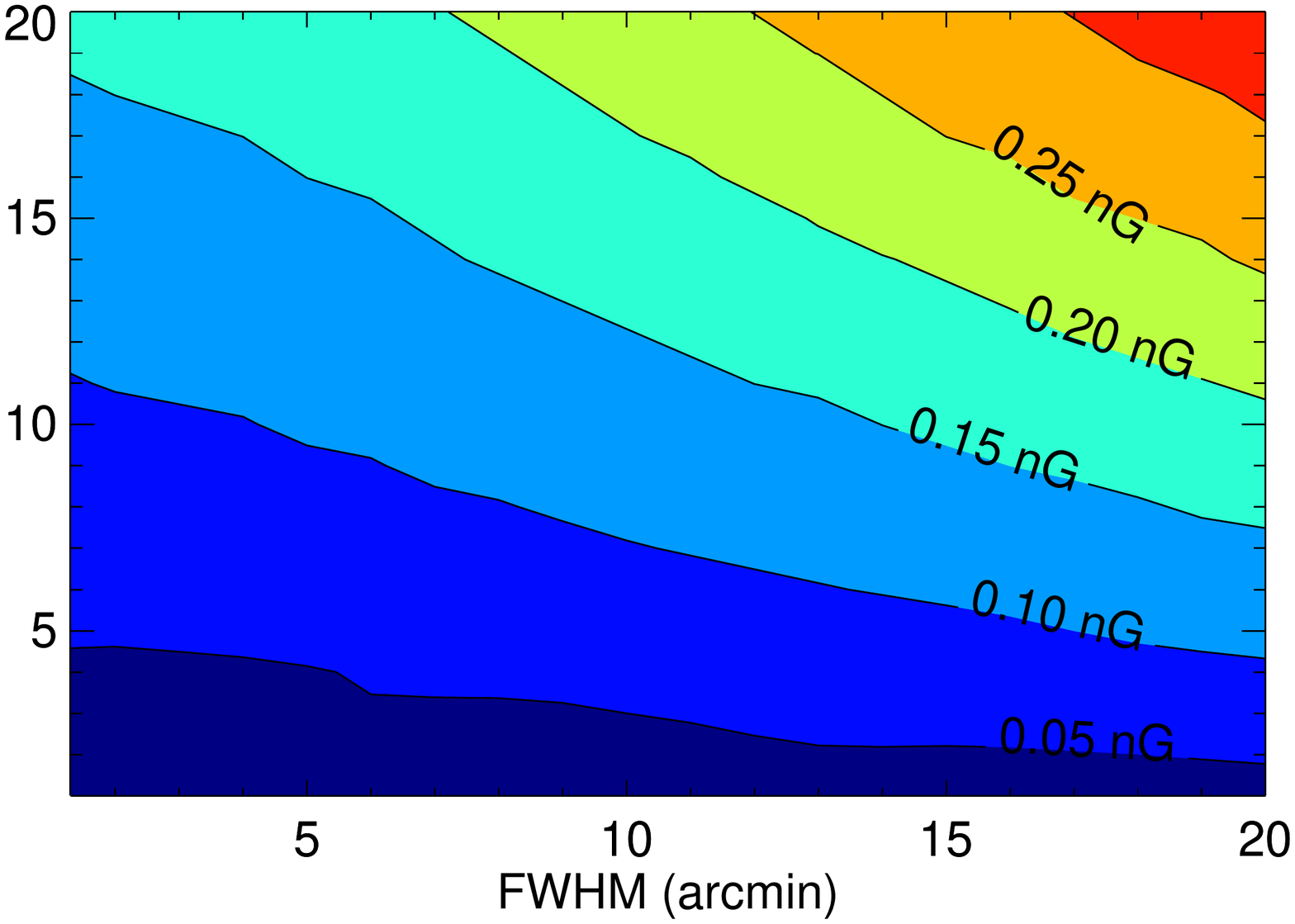}
\end{tabular}
\end{center}
\vskip -1 cm
\caption{
Contour plots of the $2\sigma$ bound on $B_{\rm eff}$ (in nG) in the 
instrumental noise -- CMB resolution plane, with the frequency channel
taken to be $30~{\rm GHz}$. The left panel shows the case with 
no de-lensing ($f_{\rm DL}=1$) and without subtracting the Milky Way RM
($f_{\rm DG}=1$). The middle 
panel is with $f_{\rm DL}=0.01$, while the right panel is with $f_{\rm DL}=0.01$ 
and $f_{\rm DG}=0.1$. It is assumed that up to $0.6$ of the sky is available 
and an optimal sky cut is found in each case.
}
\label{fig:contour}
\end{figure*}

The SNR of the detection of the primordial FR rotation angle is given by
\be
\left( S \over N \right)^2_{XB} = \sum_{L=1}^{L_{max}} {f_{\rm sky} (2L+1) [C_L^{\alpha \alpha,PMF}]^2 \over 2[C_L^{\alpha \alpha,PMF} + f_{\rm DG}C_L^{\alpha \alpha,G}+N^{XB}_L]^2} \ .
\label{eq:ebsnrP}
\ee
where $X$ stands for either E or T, depending on whether one uses the EB or TB 
estimator, $C_L^{\alpha \alpha,PMF}$ is the primordial contribution, while 
$f_{\rm DG}$ is the ``de-galaxing'' factor, {\it i.e.} the fraction of the galactic rotation 
angle spectrum that is known from other sources and can be subtracted. 

As in the case of the galactic RM, the SNR receives a non-zero contribution from all multipoles $L$, including the smallest, even for a small sky coverage. In the case when a large fraction of the sky is available, there is an optimal cut of the sky, 
$f^{\rm opt}_{\rm sky}$, for which the S/N is maximal. This is because the galactic RM contribution is weaker when large parts of the sky are cut, but 
there are more available modes when more of the sky is kept. In what follows, 
we will state when values of $f^{\rm opt}_{\rm sky}$ are different from 
$f_{\rm sky}$.

In Fig.~\ref{fig:dsnpmg}, we plot contributions per $\ln L$ to the net SNR in detection 
of the FR from a scale-invariant PMF with $B_{\rm eff}=0.1$ nG for the same two 
hypothetical experiments considered in Fig.~\ref{fig:dsngal}. In addition to the four 
cases considered in Fig.~\ref{fig:dsngal}, the green dot-dash line shows the case 
when the galactic contribution is partially subtracted for the space-based probe 
with $f_{\rm DG}=0.1$. A few observations can be made from this plot. First is that both 
experiments can detect a $0.1$ nG PMF at high significance, which constitutes a big 
improvement on prospects of detecting a PMF in CMB via non-FR signatures \cite{Ade:2013lta,Paoletti:2008ck,Paoletti:2012bb}. 
Second is that de-lensing makes a 
significant difference for both the space-based and sub-orbital experiments. Third is 
that subtracting the galactic RM moderately improves the detection at multipoles in the $4\lesssim L \lesssim 70$ range. 
Finally, we can see that most of the information comes from $4<L<200$, meaning 
that future CMB experiments can reconstruct FR maps corresponding to PMF of 
$0.1$ nG strength up to a degree resolution.

In Table~\ref{table:pmf} we present our forecasts for $2\sigma$ bounds on $B_{\rm eff}$ that can be expected from various ongoing and future CMB experiments. We present results obtained from the EB estimator, because they lead to the strongest bounds, except for the case of Planck's $30$ GHz channel, for which the TB estimator gives better constraints. For each experiment, we check the effect of de-lensing (+DL) with $f_{\rm DL}=0.01$ as well as further partial subtraction of the galactic RM (+DL+DG) with $f_{\rm DG}=0.1$. The optimal sky cuts are indicated when relevant. One of the interesting facts one can extract from this table is that very competitive bounds can be placed by POLARBEAR and QUIET, while even ${\cal O}(10^{-11}{\rm G})$ fields can be constrained in principle by future experiments.

To show the dependence of the PMF bound on the instrumental noise and resolution, we plot contours of constant $2\sigma$ bounds on $B_{\rm eff}$ in Fig.~\ref{fig:contour}. The three panels show the cases with and without de-lensing and with an additional partial subtraction of the galactic RM. An optimal sky cut was used for each set of parameters.

\section{Summary and Outlook}
\label{sec:summary}

We have studied the detectability of a scale-invariant PMF due to FR of the
CMB using polarization correlators in the presence of the foreground 
imposed
by the Milky Way magnetic field. We have found that the galactic RM is not 
a serious barrier to observe a scale-invariant PMF and that the EB quadratic
correlator is in general more powerful than the BB corelator.

We have also studied the detectability of the PMF as a function of the
sky coverage. We find that suborbital experiments can be almost as effective 
as space borne experiments, and the obstruction caused by the galaxy is relatively weak
if the observed patch is near the poles. 
Also, as the mode-coupling correlations of CMB are mostly sourced by 
the largest scale features (low $L$) of the rotation measure, a full sky
CMB map is not necessary to access the PMF on the largest scales.

Our results on the sensitivities of various experiments to the galactic 
magnetic field are summarized in Table I. Similarly in Table II we show
the effective magnetic field strength that can be constrained by those
experiments, taking into account the lensing of the CMB and the FR due
to the Milky Way. The dependence of the observable PMF to the
parameters of the experiment are shown in Fig.~\ref{fig:contour}.

Here we have focussed on the observability of CMB FR at a single frequency. 
Cross-correlating polarization maps at multiple frequencies with comparable sensitivity to FR can further boost the significance of detection. We will address this possibility in a future publication.
Also, frequency dependence of polarization can be used to separate out
FR induced B-modes from those induced by gravitational waves.

\acknowledgments We thank the two anonymous referees for their careful reading of the initially submitted version and constructive suggestions that helped to improve the paper. We are grateful to Niels Oppermann and collaborators \cite{Oppermann:2011td} for making their rotation measure maps publicly available at \cite{RMmaps}. Some of the results in this paper have been derived using the HEALPix \cite{healpix,Gorski:2004by} package. LP and TV benefited from previous collaborations with Amit Yadav. We thank Steven Gratton, Yin-Zhe Ma, Phil Mauskopf and Vlad Stolyarov for discussions and helpful comments. SD is supported by a SESE Exploration postdoctoral fellowship; LP is supported 
by an NSERC Discovery grant; TV is supported by the DOE at Arizona State University.

\end{document}